\documentclass[journal]{IEEEtran}
\usepackage{cite}
\usepackage{xcolor}

\ifCLASSINFOpdf
\else
\fi
\usepackage{soul}

\usepackage{physics}

\providecommand*{\eu}%
{\ensuremath{\mathrm{e}}}
\providecommand*{\iu}%
{\ensuremath{\mathrm{j}}}

\usepackage{amsfonts}

\allowdisplaybreaks

\providecommand*{\unit}[1]{%
    \ensuremath{\mathrm{\,#1}}} 
    
\usepackage{float}

\providecommand*{\fig}{%
    Fig.}

\usepackage{bm}  

\usepackage{graphicx}

\usepackage[version=3]{mhchem}

\begin{document}
%
\title{The effect of complex dispersion and characteristic impedance on the gain of superconducting traveling-wave kinetic inductance parametric amplifiers}
%
%
%

\author{Javier~Carrasco, Daniel~Valenzuela, Claudio~Falcón,            Ricardo~Finger, and~F.~Patricio~Mena
\thanks{J.~Carrasco is with the Electrical Engineering Department and the Department of Physics, Faculty of Physical and Mathematical Sciences, University of Chile, Santiago, Chile.}%
\thanks{D.~Valenzuela is with the Electrical Engineering Department, Faculty of Physical and Mathematical Sciences, University of Chile, Santiago, Chile.}%
\thanks{C.~Falcón is with the Department of Physics, Faculty of Physical and Mathematical Sciences, University of Chile, Santiago, Chile.}%
\thanks{R.~Finger is with the Department of Astronomy, Faculty of Physical and Mathematical Sciences, University of Chile, Santiago, Chile.}%
\thanks{F.~P.~Mena is with the National Radio Astronomy Observatory, Charlottesville, VA, USA.}%
\thanks{Contact e-mail: javier.carrasco@ug.uchile.cl.}
}

\maketitle

\begin{abstract}
Superconducting traveling-wave parametric amplifiers are a promising amplification technology suitable for applications in submillimeter astronomy. Their implementation relies on the use of Floquet transmission lines in order to create strong stopbands to suppress undesired harmonics. In the design process, amplitude equations are used to predict their gain, operation frequency, and bandwidth. However, usual amplitude equations do not take into account the real and imaginary parts of the dispersion and characteristic impedance that results from the use of Floquet lines, hindering reliable design. In order to overcome this limitation, we have used the multiple-scales method to include those effects. We demonstrate that complex dispersion and characteristic impedance have a stark effect on the transmission line's gain, even suppressing it completely in certain cases. The equations presented here can, thus, guide to a better design and understanding of the properties of this kind of amplifiers.
\end{abstract}


\IEEEpeerreviewmaketitle

\section{Introduction}
%
%
%
%

\IEEEPARstart{A}{chieving} larger bandwidths at the RF and IF bands, and improving receiver sensitivity are major challenges for future millimeter and submillimeter heterodyne observations~\cite{Carpenter2020}. As part of this effort, extensive work is being done in order to improve the performance of SIS mixers~\cite{Kojima2020} and HEMT amplifiers~\cite{Yagoubov2020}, the key components of current state-of-the-art receivers.
However, on the one hand, it is not clear if HEMT amplifiers can be further improved notwithstanding the extensive work made in understanding the reasons that limit noise temperature and operational bandwidth~\cite{Pospieszalski2017, Pospieszalski2018,Ardizzi2022}. On the other hand, even if SIS mixers are improved, connecting them to HEMT amplifiers will necessarily limit their performance~\cite{Pospieszalski2021,Kerr2014}. Recently, a promising superconducting technology that could overcome these problems has emerged \cite{Eom2012}. It uses the kinetic inductance (KI) of superconductors \cite{Zmuidzinas2012, Alexandrov2003} to produce parametric amplification in a long transmission line (TL). Devices working with this principle are dubbed Traveling-Wave Kinetic-Inductance Parametric Amplifiers (TKIPAs) \cite{Eom2012, Chaudhuri2017, Zhao2019, Malnou2021, Shu2021, Malnou2021-arxiv}.

The KI, originated by the inertia of Cooper pairs \cite{Zmuidzinas2012} in superconductors, modifies the wave-equation on the TL by adding nonlinearities which allow the mixing of wave amplitudes when more than one monochromatic wave are injected \cite{Chaudhuri2015}. Hence, it is possible to amplify the input signal if other signals, called pumps, are simultaneously injected. Nonetheless, more signals, including undesired harmonics, are also generated, compromising the amplification process. Eom et al. \cite{Eom2012} solved this problem by implementing a suitable Floquet TL, also known as \emph{dispersion engineered TL}, conformed by a periodically repeating unit cell that creates stopbands and, thus, avoids the propagation of the main undesired harmonics of the pump signal. Such a solution, however, translates into a TL with more intricate properties, namely a complex dispersion and characteristic impedance, i.e. with real and imaginary parts, that, moreover, have strong frequency dependencies, particularly close to the stopbands.

In order to design TKIPAs, a nonlinear wave equation must be solved. This is usually done by approximating the process of amplitude gain as a dynamical evolution occurring at a much larger length scale than the wavelength of the involved signals. Within this approximation, but without taking into account the complex nature of the Floquet TL, a set of nonlinear amplitude equations can be obtained \cite{Powers2017, Moloney2018}. In order to account for losses, an attempt to introduce a complex propagation constant in this approximation has been reported \cite{Zhao2019} but lacks justification when dealing with the wave behavior near stopbands.

We have included the complex nature of the Floquet TL into the amplitude equations by formally solving the nonlinear wave equation via a multiple-scales method, widely used in nonlinear physics and especially useful in traveling-wave equations \cite{Moloney2018, Nayfeh1995ch1, Nayfeh1995ch5}. We demonstrate that the use of this type of line has a profound effect on the attainable gain, in particular when the pump signal is close to a stopband. Depending on the specific properties of the used Floquet TL and the amplitude and frequency of the pump signal, our equations depart notably from the predictions given by the traditional amplitude equations.

\section{Amplitude Equations with Complex Characteristic Impedance and Dispersion}
In a TL, the electric current $I=I(z,t)$ and voltage $V=V(z,t)$ dynamics are given by the \emph{telegraph equations},

\begin{subequations}\label{eq:telegrapher}
\begin{align}
    \pdv{V}{z} &= -L \pdv{I}{t} - RI, \\
    \pdv{I}{z} &= -C \pdv{V}{t} - GV,
\end{align}
\end{subequations}
where $z$ is the position along the TL and $t$ is the time. Here, $R$ is the resistance per unit length due to losses in the conductors, $C$ is the capacitance per unit length due to the close proximity between conductors, $G$ is the conductance per unit length due to losses in the dielectric material between conductors, and $L$ is the total self-inductance per unit length between the conductors \cite{Pozar2012}. However, for TLs made out of superconductors, the inductance per unit length is a function of the current and can be modeled as
\begin{align}
    \label{eq:inductance_current}
    L = L_0 \left( 1 + \alpha_{*}\frac{I^2}{I_{*}'^2} \right),
\end{align}
where $L_0$ is the total inductance per unit length of the TL at null electric current, $\alpha_{*}$ is the ratio of kinetic inductance to total inductance, and $I_{*}'$ is a parameter comparable to the critical current $I_c$~\cite{Eom2012} (which is in the order of a few milliamperes for realizable devices). Importantly, $I_{*}=I_{*}'/\sqrt{\alpha_{*}}$ determines the strength of the nonlinear effect~\cite{Chaudhuri2015, Eom2012}.

Expression (\ref{eq:inductance_current}) comes from the fact that the current has a cubic dependence on the velocity of Cooper pairs \cite{Cho1997, Anlage1989}. It is valid at temperatures $T \ll T_c$ \cite{Anlage1989}, where $T_c$ is the critical temperature of the superconductor, commonly of the order of a few kelvins~\cite{Alexandrov2003}.

From (\ref{eq:telegrapher}) and (\ref{eq:inductance_current}), a nonlinear wave equation for the current can be derived,
\begin{align}
    \nonumber
    \left( \pdv[2]{z} - CL_0 \pdv[2]{t} - (CR + GL_0) \pdv{t} - RG \right) I\\
    \label{eq:non-lin-current}
    = \frac{L_0}{3 I_{*}^2} \left( G \pdv{t} + C \pdv[2]{t} \right) I^3.
\end{align}
It can be compactly rewritten as
\begin{align}
    \label{eq:non-lin-current_operatorsForm}
    \mathcal{L} I = \mathcal{N} I^3,
\end{align}
where $\mathcal{L}$ and $\mathcal{N}$ are, respectively, the differential operators acting on the linear and nonlinear parts of the equation. The equation $\mathcal{L}I = 0$ corresponds to the well known linear wave equation whose general solution, for waves traveling along the TL in $+z$ direction, is
\begin{align}
    \label{eq:Ilinear}
    I_{\text{linear}} = \frac{1}{2} \sum_{n} \left( A_n \eu^{\iu \omega_n t - \gamma_n z} + \text{c.c.} \right),
\end{align}
where ``c.c.'' stands for ``complex conjugate'', $\omega_n \equiv 2\pi \nu_n$ is the angular frequency of the $n-$th tone of $I$, and $\gamma_n$ are the propagation constants that fulfill the dispersion relation
\begin{align}
    \label{eq:dispRel_linear}
    \gamma_n^2 + CL_0 \omega_n^2 - \iu \omega_n (CR + GL_0) - RG = 0.
\end{align}

\begin{figure}[!t]
    \centering
    \includegraphics[width=\linewidth]{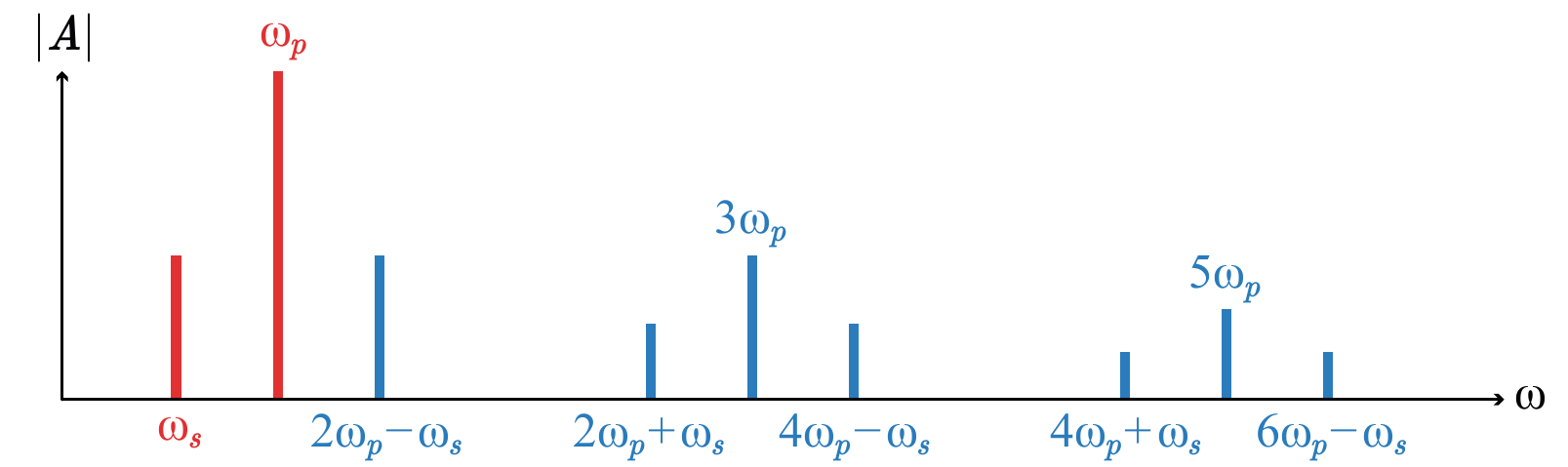}
    \caption{Fourier spectrum of the main signals relevant in the FWM process \cite{Chaudhuri2015}. This diagram shows the physically injected signals (red), and the signals that are generated in the FWM process (blue). From the latter, the idler with angular frequency $\omega_i \equiv 2\omega_p - \omega_s$, and the third harmonic of the pump with angular frequency $3\omega_p$, are the most relevant. Many more signals appear at higher frequencies, but they are less relevant because they have much smaller amplitudes.}
    \label{fig:FWM_FourierSpectrum}
\end{figure}

The solution to $\mathcal{L}I = 0$ can be obtained independently at each frequency. However, if $\alpha_{*} \neq 0$ (i.e. $\mathcal{N} \neq 0$), the term $I^3$ allows for interaction between frequencies which, under the correct conditions, can produce amplification of a target frequency signal. The level of amplification depends on a specific parameter, the \emph{pump} signal, which is an injected monochromatic wave used as source of energy in the process. This is known as \emph{parametric amplification}. Moreover, the cubic nonlinearity means that this is a four-wave-mixing (FWM) process. In consequence, the energy transfer occurs fundamentally by exchange of pairs of photons, rather than single ones \cite{Powers2017}. Therefore, in order to transfer energy from the pump to the signal, an additional monochromatic wave is required, called \emph{idler}, which also receives energy from the pump. The idler is naturally generated in the FWM process together with harmonics and sidebands \cite{Chaudhuri2015} as depicted by \fig~\ref{fig:FWM_FourierSpectrum}.

\subsection{\label{subsec:multiple_scales_method}Multiple-scales method}

We tackle the problem of solving the nonlinear current equation (\ref{eq:non-lin-current}) using a \emph{multiple-scales method} by considering the nonlinear term $\mathcal{N} I^3$ as a perturbation to the linear wave dynamics $\mathcal{L} I = 0$ which evolves at a different rate or scale~\cite{Moloney2018, Young1991}.

The method is applied to Eq. (\ref{eq:non-lin-current_operatorsForm}) rewritten in terms of $\tilde{I} \equiv I/I_{*}$. Since $|I|<I_c$ and $I_{*}\geq I_c$, then $|\tilde{I}| \equiv |I/I_{*}| < 1$, and it follows that $|\tilde{I}|^3 < |\tilde{I}|$. In consequence, the nonlinear terms can be considered perturbations to the linear equation of the current, acting at scales in $z$ larger than the evolution of the linear equation. The typical spatial scale of the linear part of (\ref{eq:non-lin-current_operatorsForm}) is the wavelength $2\pi/\beta_n$, where $\beta_n \equiv \Im{\gamma_n}$ is the wavenumber, whereas the (nonlinear) typical length scale of the envelope $A_n$ is much larger, as it will be clear later in subsection \ref{subsec:sol_nonlinWaveEq}. This information is considered by giving the amplitudes $A_n$ in (\ref{eq:Ilinear}) a dependence on $z$ but only at large scales. Then, to solve at first order, only the dominant $z$-scale affecting $A_n$ is considered, giving a new wave equation, balanced at first order. The solution to this equation must not contain singular terms, condition that results in a set of amplitude equations.

We focus on a zone close to the stopbands in Floquet TLs, commonly used to implement amplifiers using this principle \cite{Eom2012, Chaudhuri2015}. The set of amplitude equations display solutions that depend on the specific TL used to transport the signals. In particular, we obtain a new model valid in a frequency zone where $CR\omega\tilde{I}, GL_0\omega\tilde{I}, RG\tilde{I},$ and $CL_0\tilde{I}^3 \omega^2/3$ are of around the same order of magnitude because $\tilde{I}^2 \approx \frac{3G}{C\omega}$. Additionally, in the zone of interest, these terms are small enough to be considered first order perturbations of the $(\gamma^2 + \omega^2 CL_0)\tilde{I} = 0$ equation. Since this zone happens to be near the stopbands in our Floquet TLs, we call our new model \textbf{Near-Stopband Complex Transmission Line (NS-CTL)}.

In order to exemplify the results obtained within this model, we present the case $\tilde{I} = 0.2$ which is a reasonable value of operation and permits comparison with state-of-the-art devices \cite{Eom2012, Chaudhuri2015, Chaudhuri2017, Zhao2019, Shu2021}. A thorough study on the effect of varying $\tilde{I}$, modifying the validity zone will be presented elsewhere.

\subsection{\label{subsec:sol_nonlinWaveEq}Solution to the nonlinear wave equation}

After balancing (\ref{eq:non-lin-current}) at first order in the frequency zone of interest, i.e. near the stopbands, we derive the NS-CTL model,
\begin{align}
    \nonumber
    \left( \pdv[2]{z} - CL_0 \pdv[2]{t} \right) \tilde{I} &= \epsilon RG \tilde{I} + \epsilon (CR + GL_0) \pdv{t} \tilde{I} \\
    \label{eq:New3_model}
    &+ \epsilon \left( \frac{L_0 C}{3} \pdv[2]{t} \right) \tilde{I}^3.
\end{align}
This equation can be compactly expressed as
\begin{align}
    \label{eq:ap_models_eq_compact}
    \mathcal{L}^{(0)} \tilde{I} = \epsilon \left( \mathcal{L}^{(1)} \tilde{I} + \mathcal{N}^{(1)} \tilde{I}^3 \right),
\end{align}
where the positive number $\epsilon\ll1$ is explicitly written to identify the perturbative first order terms. $\mathcal{L}^{(0)}$ is the operator acting on the unperturbed linear equation, while $\mathcal{L}^{(1)}$, and $\mathcal{N}^{(1)}$ are the operators acting as first order perturbations on the linear and nonlinear parts of the equation, respectively.

Very far from the stopbands in a Floquet TL, the right-hand-side of (\ref{eq:ap_models_eq_compact}) reduces to $\epsilon \mathcal{N}^{(1)} \tilde{I}^3$ because the $RG, CR$ and $GL_0$ quantities are too small in those cases, making the term $\mathcal{L}^{(1)} \tilde{I}$ to act at orders smaller than $\epsilon$. However, as we get closer to the stopbands, $RG, CR$ and $GL_0$ increase in magnitude, and the term $\mathcal{L}^{(1)} \tilde{I}$ must be included as done in (\ref{eq:ap_models_eq_compact}). We operate our TKIPA in the frequency zone where this is justified.

Within the NS-CTL model, after applying the multiple-scales method, we obtain the \emph{amplitude equation} for the $m-$th signal,
\begin{align}
    \nonumber
    \pdv{A_m}{z} &= \iu g_m A_m - 2 \alpha_m A_m \\
    \label{eq:ampEq_generalForm}
    &+ \frac{\iu f_m \eu^{\iu \beta_m z}}{3 \cdot 8 \cdot I_{*}^2} \braket{\eu^{\iu \omega_m t}}{\left( \sum_n^N A_n \eu^{\iu (\omega_n t - \beta_n) z} + \text{c.c.} \right)^3},
\end{align}
where
\begin{align}
    \label{eq:f_factor}
    f_m &= \frac{1}{2\beta_m} \left( \alpha_m^2 - \beta_m^2 - \frac{\abs{\gamma_m}^2}{\abs{\eta_m}^2} (r_m^2 - x_m^2) \right),\\
    g_m &= \frac{\alpha_m^2 r_m^2 - \beta_m^2 x_m^2}{\beta_m (r_m^2 + x_m^2)},
\end{align}
and
\begin{align}
    \gamma_m &= \sqrt{ (R_m + \iu \omega_m L_{0,m}) (G_m + \iu \omega_m C_m) },\\
    \eta_m &= \sqrt{ \frac{R_m + \iu \omega_m L_{0,m}}{G_m + \iu \omega_m C_m} }.
\end{align}
In Floquet TLs the propagation constant $\gamma_m \equiv \alpha_m + \iu \beta_m$ and the characteristic impedance $\eta_m \equiv r_m + \iu x_m$ are strongly dependent of the frequency in the zone of interest, and have non-negligible real and imaginary parts. We have used the Hermitian inner product defined by
\begin{align}
    \braket{a(t)}{b(t)} \equiv \frac{1}{2\pi} \int_0^{2\pi} a^{*}(t) b(t) \dd{t},
\end{align}
for any couple of time-dependent $2\pi$-periodic complex valued continuous functions $a(t), b(t)$ as a projector, which allows us to quantify the contribution of one wave to another.

From (\ref{eq:ampEq_generalForm}), we observe the two terms contributing to the oscillatory part of the solution for $A_m(z)$ scale in space as $\abs{2\pi/g_m}$ and $\abs{2\pi\cdot3\cdot8/f_m}$, respectively. Therefore, in frequency zones where $\alpha_m \ll \beta_m$ and $\abs{x_m} \ll r_m$, the minimum spatial scale of $A_m(z)$ is given by $\abs{2\pi\cdot3\cdot8/f_m} \approx 2\pi\cdot24/\beta_m$. This scale is 24 times larger than the wavelength $2\pi/\beta_m$, justifying the use of the multiple-scales method. The exact scale is $\min\{\abs{2\pi/g_m}, \abs{2\pi\cdot24/f_m}\}$, which varies with the frequency. The scale given by $\pi/\alpha_m$ is just the decay in $z$ 
of the amplitudes due to losses, which is much larger than all other scales, since $\alpha_m \ll \beta_m$ for all frequencies outside stopbands.

The \textbf{Lossless Transmission Line (LTL)} model used in the literature \cite{Eom2012} corresponds to the NS-CTL model but changing the value of the TL parameters ($R_m, G_m, C_m, L_{0,m}$) for ones that neglect $x_m \equiv \Im{\eta_m}$ and $\alpha_m \equiv \Re{\gamma_m}$. The NS-CTL model reduces to the LTL model in the limiting case of $x_m=\alpha_m=0$, because then $R_mG_m=C_mR_m+G_mL_{0,m}=0$. This makes $g_m=0$ and $f_m=-\beta_m$, which do not hold for frequencies near the stopbands.

To make the amplitude equations (\ref{eq:ampEq_generalForm}) explicit, we need to set the number of waves, $N$. The signal ($s$), pump ($p$), and idler ($i$) waves are the minimum required to describe the FWM dynamics and obtain parametric amplification.

\subsection{\label{subsec:sip}Signal, pump, and idler interaction ($s\text{-}p\text{-}i$)}

By setting $N=3$ in the amplitude equations (\ref{eq:ampEq_generalForm}) with $n,m=s,p,i$, we obtain a set of three amplitude equations,
\begin{subequations}\label{eq:ampEq_spi}
\begin{align}
    \nonumber
    \pdv{A_s}{z} &= \iu g_s A_s - 2 \alpha_s A_s + \iu \frac{f_s}{8 I_{*}^2} \\
    \label{eq:ampEq_spi_s}
    &\times \left[ A_s ( \abs{A_s}^2 + 2 \abs{A_i}^2 + 2 \abs{A_p}^2) + A_i^{*} A_p^2 \eu^{\iu \Delta \beta z} \right], \\
    \nonumber
    \pdv{A_i}{z} &= \iu g_i A_i - 2 \alpha_i A_i + \iu \frac{f_i}{8 I_{*}^2} \\
    \label{eq:ampEq_spi_i}
    &\times \left[ A_i ( 2\abs{A_s}^2 + \abs{A_i}^2 + 2 \abs{A_p}^2) + A_s^{*} A_p^2 \eu^{\iu \Delta \beta z} \right], \\
    \nonumber
    \pdv{A_p}{z} &= \iu g_p A_p - 2 \alpha_p A_p + \iu \frac{f_p}{8 I_{*}^2} \\
    \label{eq:ampEq_spi_p}
    &\times \left[ A_p ( 2 \abs{A_s}^2 + 2 \abs{A_i}^2 + \abs{A_p}^2) + 2 A_p^{*} A_s A_i \eu^{-\iu \Delta \beta z} \right],
\end{align}
\end{subequations}
where $\Delta \beta \equiv \beta_s + \beta_i - 2 \beta_p$.

Interestingly, by writing $A_m \equiv \abs{A_m} \eu^{\iu \phi_m}$, the evolution of the modules of the amplitudes can be found as
\begin{subequations}\label{eq:ampEq_spi_MagPhase}
\begin{align}
    \pdv{\abs{A_s}}{z} &= - 2 \alpha_s \abs{A_s} - \frac{f_s}{8 I_{*}^2} \abs{A_i} \abs{A_p}^2 \sin{\Theta(z)}, \\
    \pdv{\abs{A_i}}{z} &= - 2 \alpha_i \abs{A_i} - \frac{f_i}{8 I_{*}^2} \abs{A_s} \abs{A_p}^2 \sin{\Theta(z)}, \\
    \pdv{\abs{A_p}}{z} &= - 2 \alpha_p \abs{A_p} + 2\frac{f_p}{8 I_{*}^2} \abs{A_s} \abs{A_i} \abs{A_p} \sin{\Theta(z)}, \\ \nonumber
    \pdv{\Theta}{z} &= \Delta \beta - \Delta g + \frac{1}{8 I_{*}^2} \Biggl[ \abs{A_s}^2 (f_s - 2 \Delta f)
    \\ \nonumber
    &+ \left. \abs{A_i}^2 (f_i - 2 \Delta f) - 2\abs{A_p}^2 (f_p + \Delta f) + \cos\Theta(z) \right.
    \\ \label{eq:ampEq_spi_MagPhase_Theta}
    &\times \left( 4 \abs{A_s} \abs{A_i} f_p - \frac{\abs{A_i}\abs{A_p}^2}{\abs{A_s}} f_s - \frac{\abs{A_s}\abs{A_p}^2}{\abs{A_i}} f_i \right) \Biggr],
\end{align}
\end{subequations}
where $\Theta(z) \equiv \Delta \beta z - \Delta \phi(z)$, $\Delta \phi(z) \equiv \phi_s + \phi_i - 2\phi_p$, $\Delta g \equiv g_s + g_i - 2g_p$, and $\Delta f \equiv f_s + f_i - 2f_p$. The phase $\Theta(z)$ is the \emph{total phase mismatch} that consists of a \emph{linear}, $\Delta \beta z$, and a \emph{nonlinear phase mismatch}, $\Delta \phi(z)$. The phase mismatch is relevant because, depending on its value, transfer of energy flows from the pump toward the signal and idler waves or viceversa \cite{Hansryd2002}. Since $f_m < 0$, the former occurs for $\Theta \in (0, \pi)$, and the latter for $\Theta \in (-\pi, 0)$. Moreover, in the very special cases of $\Theta = 0, \pm\pi$, no transfer of energy happens between the signals. Therefore, thanks to the existence of the phase mismatch, these amplitude equations predict the possibility of amplifying the target signal. Furthermore, the ideal situation occurs when $\Theta = \frac{\pi}{2}$, maximizing the transfer rate of energy toward the target signal.

\subsection{Including the pumps's third harmonic ($s\text{-}p\text{-}i\text{-}3p$)}

The $s\text{-}p\text{-}i$ case can be improved by including the wave of the third harmonic of the pump ($3p$). For doing so, we set $N=4$, implying that this time we get a set of four amplitude equations,
\begin{subequations}\label{eq:ampEq_spi3p}
\begin{align}
    \nonumber
    \pdv{A_s}{z} &= \iu g_s A_s - 2 \alpha_s A_s + \iu \frac{f_s}{8 I_{*}^2} \\
    \nonumber
    &\times \left[ A_s ( \abs{A_s}^2 + 2 \abs{A_i}^2 + 2 \abs{A_p}^2 + 2 \abs{A_{3p}}^2) \right. \\ 
    \label{eq:ampEq_spi3p_s}
    &\qquad + \left. A_i^{*} A_p^2 \eu^{\iu \Delta \beta z} + 2 A_i^{*} A_p^{*} A_{3p} \eu^{\iu \Delta \beta_2 z} \right], \\ \nonumber
    \pdv{A_i}{z} &= \iu g_i A_i - 2 \alpha_i A_i + \iu \frac{f_i}{8 I_{*}^2} \\
    \nonumber
    &\times \left[ A_i ( 2\abs{A_s}^2 + \abs{A_i}^2 + 2 \abs{A_p}^2 + 2 \abs{A_{3p}}^2) \right. \\ 
    \label{eq:ampEq_spi3p_i}
    &\qquad + \left. A_s^{*} A_p^2 \eu^{\iu \Delta \beta z} + 2 A_s^{*} A_p^{*} A_{3p} \eu^{\iu \Delta \beta_2 z} \right], \\ \nonumber
    \pdv{A_p}{z} &= \iu g_p A_p - 2 \alpha_p A_p + \iu \frac{f_p}{8 I_{*}^2} \\
    \nonumber
    &\times \left[ A_p ( 2 \abs{A_s}^2 + 2 \abs{A_i}^2 + \abs{A_p}^2 + 2 \abs{A_{3p}}^2) \right. \\ 
    \nonumber
    & + \left. 2 A_p^{*} A_s A_i \eu^{-\iu \Delta \beta z} + 2 A_s^{*} A_i^{*} A_{3p} \eu^{\iu \Delta \beta_2 z} \right. \\ 
    \label{eq:ampEq_spi3p_p}
    & + \left. A_{3p} (A_p^{*})^2 \eu^{\iu \Delta \beta_3 z} \right], \\ \nonumber
    \pdv{A_{3p}}{z} &= \iu g_{3p} A_{3p} - 2 \alpha_{3p} A_{3p} + \iu \frac{f_{3p}}{8 I_{*}^2} \\
    \nonumber
    &\times \left[ A_{3p} ( 2 \abs{A_s}^2 + 2 \abs{A_i}^2 + 2 \abs{A_p}^2 + 2 \abs{A_{3p}}^2) \right. \\ 
    \label{eq:ampEq_spi3p_3p}
    &\qquad + \left. \frac{1}{3} A_p^3 \eu^{-\iu \Delta \beta_3 z} + 2 A_s A_i A_p \eu^{-\iu \Delta \beta_2 z} \right],
\end{align}
\end{subequations}
where $\Delta \beta_2 \equiv \beta_s + \beta_i + \beta_p - \beta_{3p}$, and $\Delta \beta_3 \equiv 3\beta_p - \beta_{3p}$.

The presence of the third harmonic of the pump can greatly reduce the signal gain in the FWM process of the TKIPA. To properly account for this effect, the $3p$ wave must be taken into consideration, requiring the use of the amplitude equations~(\ref{eq:ampEq_spi3p}). Nevertheless, we know from the linear wave solution in TLs, i.e. (\ref{eq:Ilinear}) and (\ref{eq:dispRel_linear}), that the $3p$ wave is suppressed if its frequency is inside a stopband. Therefore, we expect the same to happen with the set of amplitude equations~(\ref{eq:ampEq_spi3p}), allowing to simplify it to (\ref{eq:ampEq_spi}). In section \ref{sec:ResDis} we simulate both sets of amplitude equations to verify that this is the case.


\section{\label{sec:OLDvsNEW}Differences between LTL and NS-CTL models}
The amplitude equations of the LTL model can be obtained for the $s\text{-}p\text{-}i$ and $s\text{-}p\text{-}i\text{-}3p$ cases by setting $\alpha_m = x_m = 0$, which gives $g_m=\alpha_m=0$ and $f_m=-\beta_m$ in (\ref{eq:ampEq_spi})-(\ref{eq:ampEq_spi3p}) for all $m=s,p,i,3p$. This sets two important differences between the models, one related to the factor $f_m$ and other to $g_m$.

In the NS-CTL model, the relevant constant factor for amplification is $f_m$. This fact can be seen from (\ref{eq:ampEq_spi_MagPhase}) by noticing that the coupling terms, allowing energy transfer, are proportional to $f_m$. Hence, in order to obtain gain of the target signal, the larger it can be, the better.

In the LTL model, instead, the relevant constant factor for amplification reduces to $-\beta_m$, obtained neglecting the effect of $\alpha_m$ and $x_m$ in (\ref{eq:f_factor}). Therefore, the LTL model has less degrees of freedom to optimize this factor. Nonetheless, $f_m \approx -\beta_m$ is an acceptable approximation for frequencies near stopbands where the NS-CTL model is valid.

On the other hand, the term $g_m$ in the NS-CTL model affects the equation for $\abs{A_m}$ through the phase mismatch $\Theta(z)$. This effect can be concluded by noticing the presence of the $\Delta g$ term in (\ref{eq:ampEq_spi_MagPhase_Theta}). Therefore, $g_m$ affects (\ref{eq:ampEq_spi_MagPhase}a-c) through the $\sin(\Theta)$ term, which multiplies the $f_m$ factor. Consequently, $g_m$ can produce a big impact in the amplification process by increasing or reducing the rate of energy transfer from the pump to the signal and idler. This is the most important effect that is absent in the LTL model, where $g_m=0$, as it is the only difference affecting directly the nonlinear dynamics, and therefore, amplification. Hence, if there is any relevant difference in the simulations between the LTL and NS-CTL models, it is due to the effect of $g_m$ on the phase mismatch $\Theta(z)$.


\section{\label{sec:ResDis}Results and Discussion}

We have designed and simulated a Floquet coplanar waveguide (CPW) TL made out of a central TL, with propagation constant $\gamma_0 \equiv \alpha_0 + \iu \beta_0$ and characteristic impedance $\eta_0 \equiv r_0 + \iu x_0$, that contains small sections of different strip width that are periodically repeated. The complete Floquet TL has an effective propagation constant $\gamma \equiv \alpha + \iu \beta \neq \gamma_0$ and effective characteristic impedance $\eta \equiv r + \iu x \neq \eta_0$. This line is designed to have frequency stopbands, defined by $\delta \alpha \equiv \alpha - \alpha_0 \neq 0$, and a nonlinear dispersion relation $\beta(\nu)$ near the stopbands, identified by $\delta \beta \equiv \beta - \beta_0 \neq 0$. The latter is very important, because it allows to tune the linear phase mismatch $\Delta \beta z \equiv \beta_s z + \beta_i z - 2\beta_p z$, relevant for the amplification of the target signal, as discussed in subsection~\ref{subsec:sip}.

\subsection{Design}

\begin{figure}[!t]
    \centering
    \includegraphics[width=\linewidth]{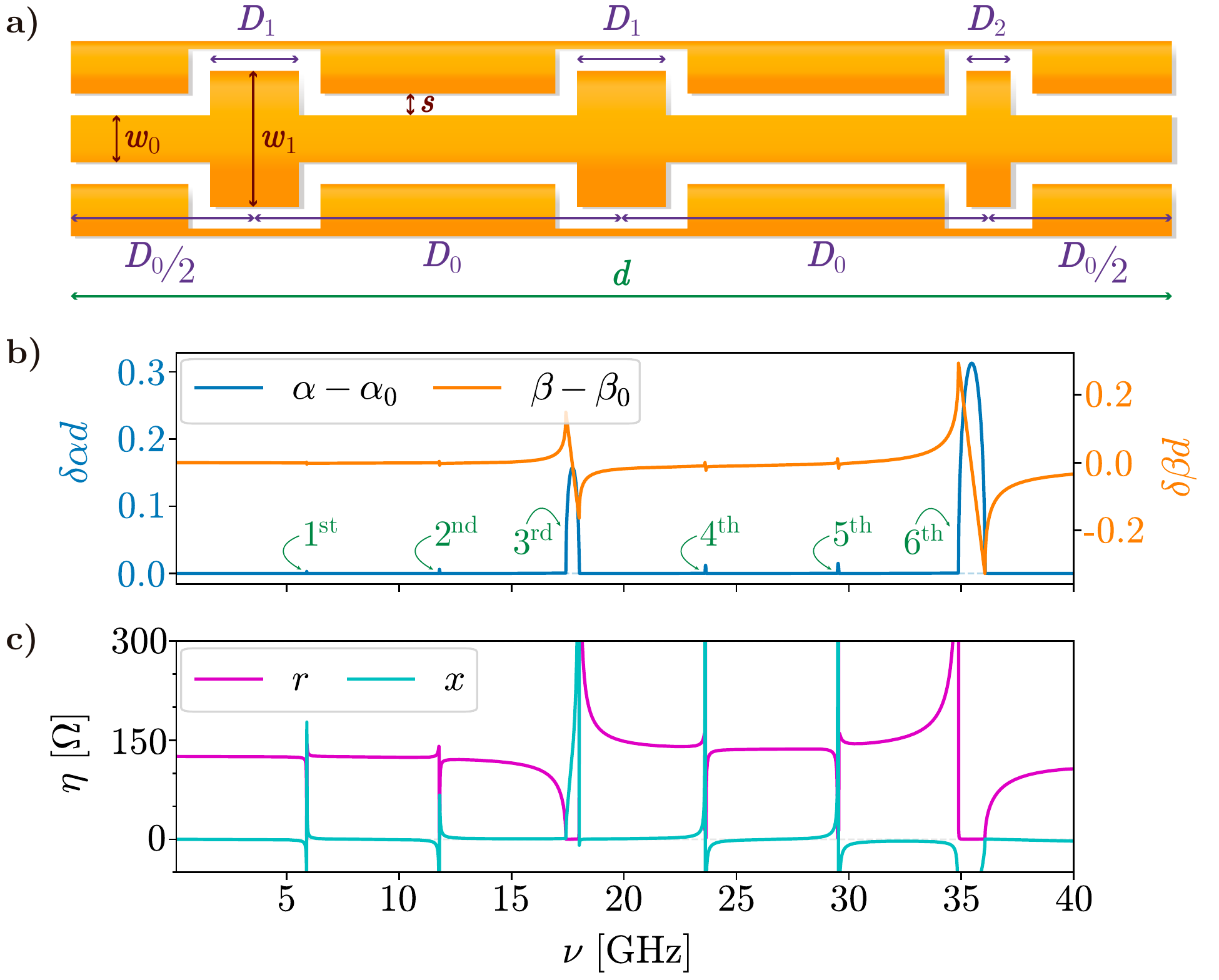}
    \caption{(a) CPW-based Floquet transmission line consisting of a central line of width $w_0$ and dispersive loads of width~$w_1$, both with the same thickness~$t$. Their values, together with those of the separation $s$ and the lengths $D_0, D_1, D_2$ and $d$, are given in Table \ref{tab:unitCellvals_CPW}. (b) Dispersion of the designed Floquet TL. Six stopbands are shown and enumerated in green. They are determined by $\delta \alpha \equiv \alpha - \alpha_0 \neq 0$, where the subindex ``0'' indicates ``central line''. Around the stopbands a steeply changing nonlinear dispersion relation is observed, i.e. $\delta \beta \equiv \beta - \beta_0 \neq 0$.
    (c) Characteristic impedance of the Floquet TL. It is noticeable that $r$ and $x$ have a strong variation close to the stopbands.}
    \label{fig:res_CPW_design}
\end{figure}

Our Floquet TL design is shown in \fig~\ref{fig:res_CPW_design}, where we have used a CPW geometry. The design parameters are given in Tables \ref{tab:simParams_CPW} and \ref{tab:unitCellvals_CPW}. The parameters $\alpha_0$ and $x_0$ are negligible compared to $\beta_0 / \nu \approx 112 \unit{m}^{-1}\unit{GHz}^{-1}$ and $r_0 \approx 127\ \Omega$, respectively. The parameters were calculated for a temperature $T = 4$ K. The superconductor's conductivity was calculated via the formulation given by Mattis and Bardeen~\cite{MattisBardeen1958, Zhao2017}. The superficial impedance and geometric factors of the CPW were calculated using the method described by Zhao et al.~\cite{Zhao2018}. Table~\ref{tab:simParams_CPW} lists the material properties of the dielectric and superconductor making up the designed CPW line.

\begin{figure}[!htb]
    \centering
    \includegraphics[width=\linewidth]{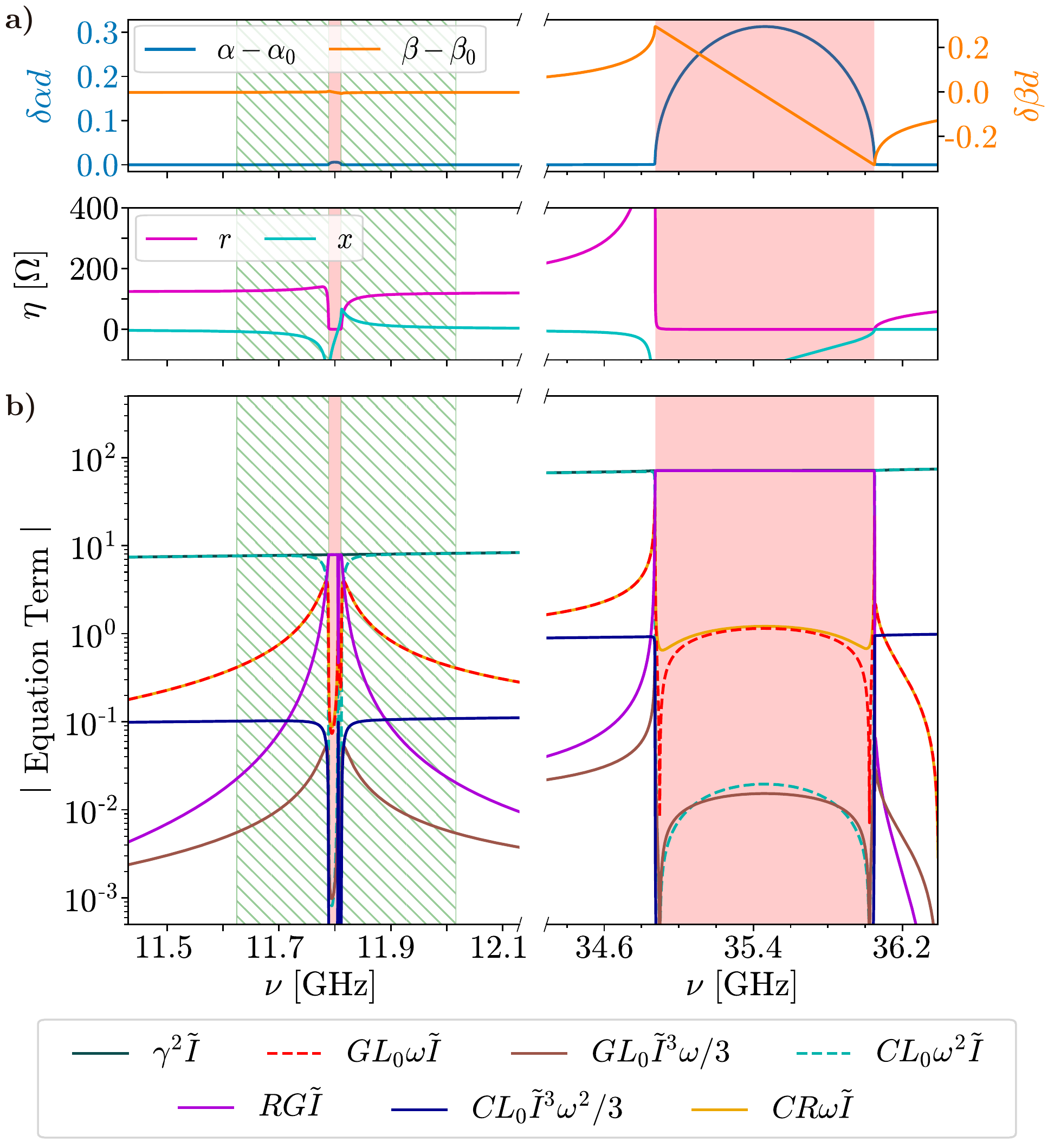}
    \caption{(a) Dispersion relation and characteristic impedance from Fig. \ref{fig:res_CPW_design}, zoomed in around the second (left column) and sixth (right column) stopbands (marked as red zones). (b) Magnitudes of the different terms of (\ref{eq:non-lin-current}) for the case $\tilde{I} \equiv I/I_{*} = 0.2$. The green dashed zones on the left, correspond to frequencies where its third harmonic $3\nu$ falls within the stopband shown on the right.}
    \label{fig:res_CPW_sb2_designZoom}
\end{figure}

In our simulations, we have set the operation of the TKIPA with the pump near the second stopband and its third harmonic in the sixth stopband. The relevant quantities at these frequencies are shown in \fig~\ref{fig:res_CPW_sb2_designZoom}, where the order of magnitude of each of the terms of (\ref{eq:non-lin-current}) are plotted as functions of the frequency. Our NS-CTL model is valid in a frequency zone where the terms $CR\omega\tilde{I}, GL_0\omega\tilde{I}, RG\tilde{I},$ and $CL_0\tilde{I}^3 \omega^2/3$ are of around the same order $\epsilon$ below the dominant term $(\gamma^2 + \omega^2 CL_0)\tilde{I}$. This happens at frequencies near any of the stopbands, including our choice. We choose the pump frequency $\nu_p$ to be in the green dashed regions because then the $3p$ wave falls inside the sixth stopband allowing its suppression.

\begin{table}[!t]
    \centering
    \caption{Parameters of the materials making up the CPW line.}
    \begin{tabular}{ccccccc}
        \hline
        \multicolumn{1}{c}{\rule{0pt}{0ex}} & \multicolumn{1}{c}{\rule{0pt}{0ex}} & \multicolumn{2}{c}{\rule{0pt}{0ex}} & \multicolumn{1}{c}{\rule{0pt}{0ex}} & \multicolumn{2}{c}{\rule{0pt}{0ex}}\\[-3pt]
         & & \multicolumn{2}{c}{\ce{NbTiN}} & & \multicolumn{2}{c}{\ce{Si}}\\[5pt]
        \hline\rule{0pt}{3ex}
        Parameter & & $T_c$ & $\rho_N$ & & $\epsilon_r$ & $\tan\delta$ \\
        (Unit) & & (K) & ($\mu\Omega\cdot$ cm) & & -- & -- \\[5pt]
        \hline\rule{0pt}{3ex} 
        Value & & 14.7 & 100 & & 11.44 & 0 \\[5pt]
        Reference & & \cite{Jiang2009, Thoen2017} & \cite{Eom2012} & & \cite{Lamb1996} & -- \\[5pt]
        \hline
    \end{tabular}
    \label{tab:simParams_CPW}
\end{table}

\begin{table}[!t]
    \centering
    \caption{Geometric parameters of the CPW Floquet transmission line.}
    \begin{tabular}{cccccccc}
        \hline\rule{0pt}{3ex}
        $t$ & $s$ & $w_0$  & $w_1$ & $D_0$ & $D_1$ & $D_2$ & $d$ \\
        (nm) & ($\mu$m) & ($\mu$m) & ($\mu$m) & (mm) & ($\mu$m) & ($\mu$m) & (mm) \\[5pt]
        \hline\rule{0pt}{3ex}
        35 & 1 & 1 & 3.4 & 1.578 & 60 & 50 & 4.734 \\[5pt]
        \hline
    \end{tabular}
    \label{tab:unitCellvals_CPW}
\end{table}

\subsection{Effect of the pump's third harmonic}
In order to study the effect of the pump's third harmonic and check the consistency of (\ref{eq:ampEq_spi}) and~(\ref{eq:ampEq_spi3p}), we applied them to two different pump frequencies, $\nu_p$. Furthermore, we repeated the simulations using the corresponding approximations for the traditional LTL model.

\begin{figure}[!t]
    \centering
    \includegraphics[width=\linewidth]{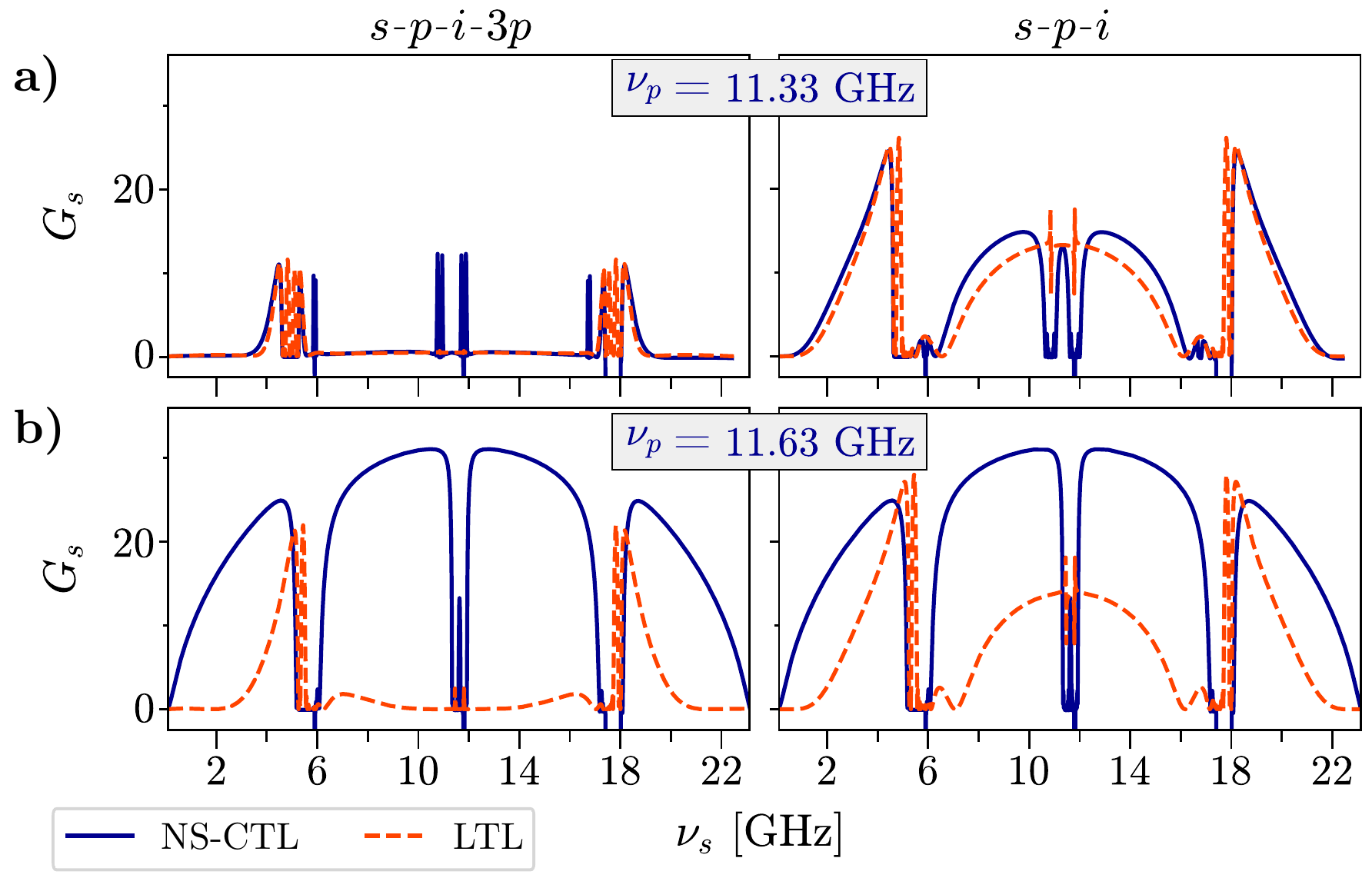}
    \caption{Gain of the target signal after traveling $z/d=150$ unit cells through the parametric amplifier. The simulations correspond to two different values of pump frequency $\nu_p$ near the second stopband with an initial pump amplitude $A_p^0 = 0.2 I_{*}$. Situations in which the third harmonic $3\nu_p$ does not fall (a) and falls (b) in a stopband are presented. The results show that only if the $3p$ signal is suppressed, it can be correctly disregarded in the NS-CTL model.}
    \label{fig:res_CPW_sb2_gains_cases}
\end{figure}

The first frequency, $\nu_p=11.33$~GHz, corresponds to a situation where $3\nu_p$ is well outside the sixth stopband. In this case, the $3p$ wave should not be suppressed and all the signal gain should be lost. The situation is well captured by (\ref{eq:ampEq_spi3p}), in both the NS-CTL and LTL models, as demonstrated by the left panel of Fig.~\ref{fig:res_CPW_sb2_gains_cases}a. On the contrary, (\ref{eq:ampEq_spi}) wrongly predicts gain as it does not consider the pump's third harmonic (see right panel of Fig.~\ref{fig:res_CPW_sb2_gains_cases}a).

In the second situation, $\nu_p=11.63$~GHz, $3\nu_p$ is well inside the sixth stopband. Here, the $3p$ wave should be suppressed and the TL should produce gain. The situation is now well described by (\ref{eq:ampEq_spi3p}) only in the NS-CTL model (left panel of Fig.~\ref{fig:res_CPW_sb2_gains_cases}b). The LTL model cannot predict gain because the assumption $\alpha=0$ impedes the detection of stopbands, meaning that the $3p$ wave is not suppressed in this simulation. Importantly now, for the NS-CTL case, (\ref{eq:ampEq_spi}) and~(\ref{eq:ampEq_spi3p}) give identical results (both panels of Fig.~\ref{fig:res_CPW_sb2_gains_cases}b), demonstrating the consistency of the framework used here.

In \fig~\ref{fig:res_CPW_sb2_gains_cases}b we also observe that, compared to the LTL model, the NS-CTL model predicts higher signal gain. The reason of this result is that the total phase mismatch $\Theta(z)$ evolves differently in each model. We thoroughly explain this effect in subsection \ref{subsec:effect_of_Theta}.

\subsection{Effect of the stopband on the gain}

\begin{figure}[!t]
    \centering
    \includegraphics[width=\linewidth]{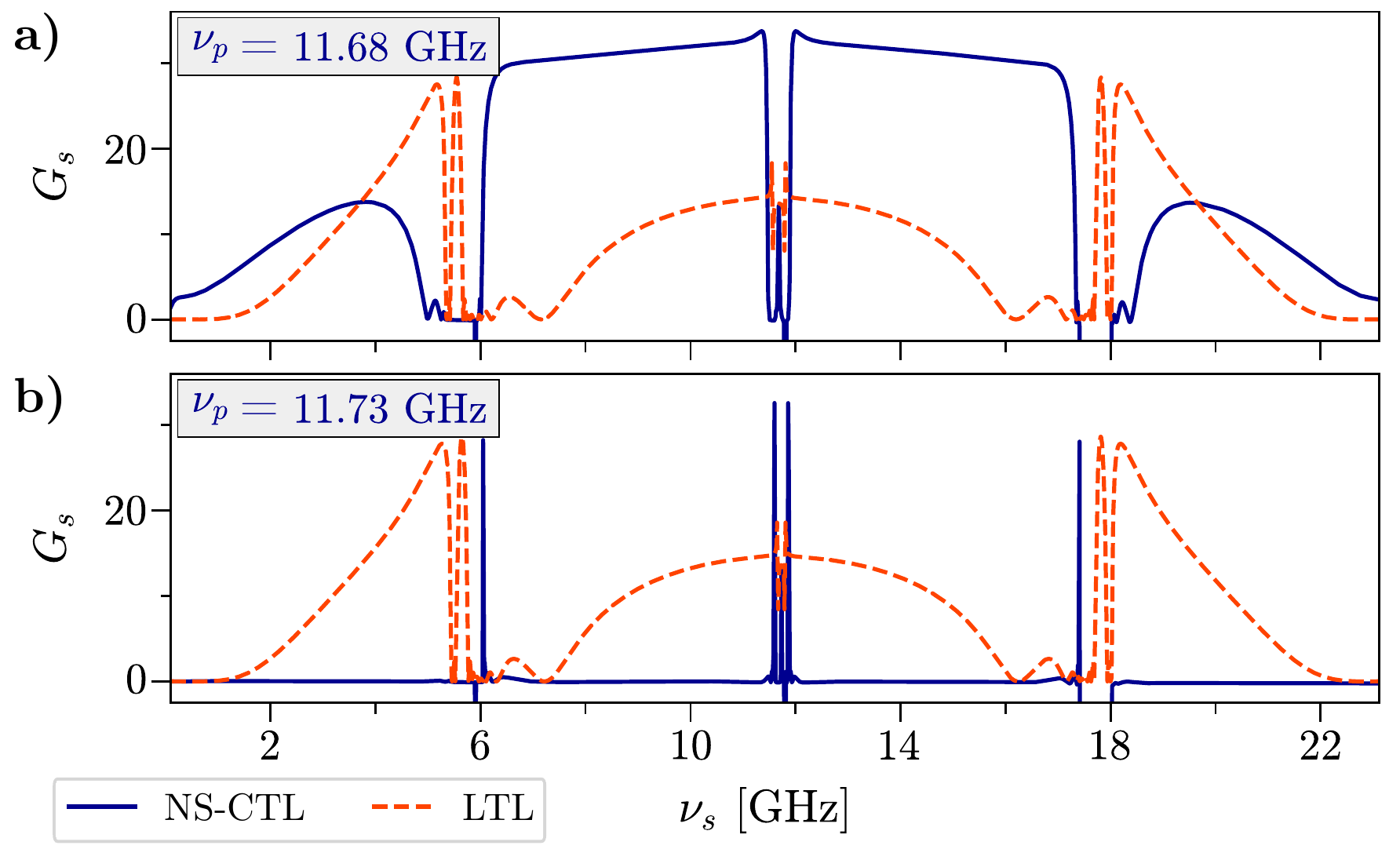}
    \caption{Parametric gain of the target signal after traveling $z/d=150$ unit cells. The simulations correspond to two values of $\nu_p$ close to the second stopband, suppressing the $3p$ signal. In both cases an initial pump amplitude $A_p^0 = 0.2 I_{*}$ was used. (a) In the first case, the NS-CTL model predicts an enhancement of the gain of the target signal. (b) When the pump frequency is even closer to the stopband, the loss term $|RG \tilde{I}|$ dominates over the nonlinear one, $|CL_0 \tilde{I}^3 \omega^2/3|$, resulting in no gain.}
    \label{fig:res_CPW_sb2_gains_cases2}
\end{figure}

Now we study the effect of moving 
$\nu_p$ closer to the stopband. With that purpose, we performed two more simulations, one with $\nu_p=11.68$~GHz (\fig~\ref{fig:res_CPW_sb2_gains_cases2}a), and another with $\nu_p=11.73$~GHz (\fig~\ref{fig:res_CPW_sb2_gains_cases2}b). In both cases the third harmonic of the pump is being suppressed, so we only show results for the $s\text{-}p\text{-}i$ case.

By comparing Figs.~\ref{fig:res_CPW_sb2_gains_cases2}a and \ref{fig:res_CPW_sb2_gains_cases}b, we observe that setting $\nu_p$ closer to a stopband results in larger gain, either using the LTL or the NS-CTL model. Nevertheless, if we keep moving $\nu_p$ even closer to the stopband, we predict no gain with the NS-CTL model (see \fig~\ref{fig:res_CPW_sb2_gains_cases2}b). This occurs because at frequencies too close to a stopband (like $\nu_p = 11.73$~GHz), the loss term $RG \tilde{I}$ dominates over the nonlinear term, $CL_0 \tilde{I}^3 \omega^2/3$. Therefore, the nonlinear dynamics responsible for amplification are overcome by loss dynamics, resulting in no gain of the target signal. The LTL model does not consider this complexity because it neglects $\alpha$ and $x$, implying neglecting $RG$. Consequently, the LTL model still predicts gain at the same frequency.

\subsection{\label{subsec:effect_of_Theta}Effect of $\Theta(z)$ on the gain}

\begin{figure}[!t]
    \centering
    \includegraphics[width=\linewidth]{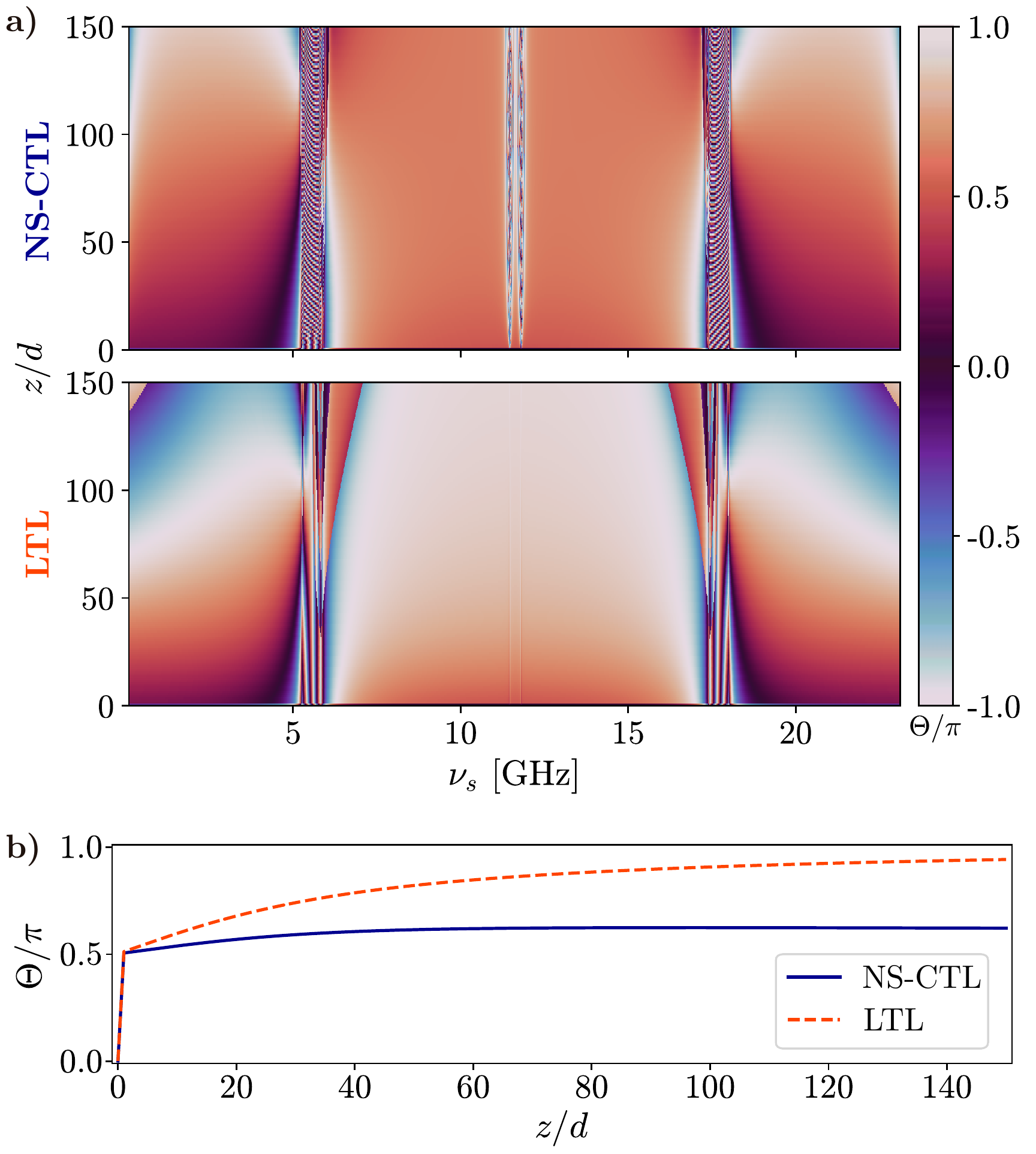}
    \caption{(a) Color-coded plot of the evolution of the phase mismatch $\Theta(z)$, for the NS-CTL and LTL models, as function of the signal frequency $\nu_s$. The simulation corresponds to the TKIPA pumped at $\nu_p=11.63$~GHz. It is observed that $\Theta(z)$ is better stabilized near $\pi/2$ along all the evolution in the NS-CTL model, meaning an energy transfer from the pump to the idler and target signals close to the optimal value along all $z$. (b) Cut of the $\Theta(z)$ evolution at $\nu_s = 10$~GHz to stress the difference between the two models.}
    \label{fig:res_CPW_sb2_theta_intPlot}
\end{figure}

As explained in section \ref{sec:OLDvsNEW}, the main difference between the NS-CTL and LTL models resides in the $g_m$ term that directly modifies the phase mismatch $\Theta(z)$. Therefore, the differences observed between the two models (like, e.g., in \fig~\ref{fig:res_CPW_sb2_gains_cases}b) should be justified by the evolution of $\Theta(z)$ in each case. To illustrate that this is indeed the case, \fig~\ref{fig:res_CPW_sb2_theta_intPlot} presents the evolution of $\Theta(z)$ as a function of the signal frequency for $\nu_p=11.63$~GHz. We see how the NS-CTL model stabilizes $\Theta(z)$ close to its optimal value $\pi/2$ along the whole dynamical evolution (i.e. through $z$). Instead, in the LTL model, $\Theta(z)$ varies between optimal and counter-optimal values, for a given signal frequency $\nu_s$, which means that at times of the evolution, the energy transfer is mainly going from the signal and idler to the pump, contrary to the desired performance. Consequently, as long as the nonlinear term $|CL_0 \tilde{I}^3 \omega^2/3|$ is larger than the loss term $|RG \tilde{I}|$, the LTL model does not predict as much gain of the signal as the NS-CTL model, explaining the difference between the models.

\subsection{Effect of the initial pump amplitude on the gain}

\begin{figure}[!t]
    \centering
    \includegraphics[width=\linewidth]{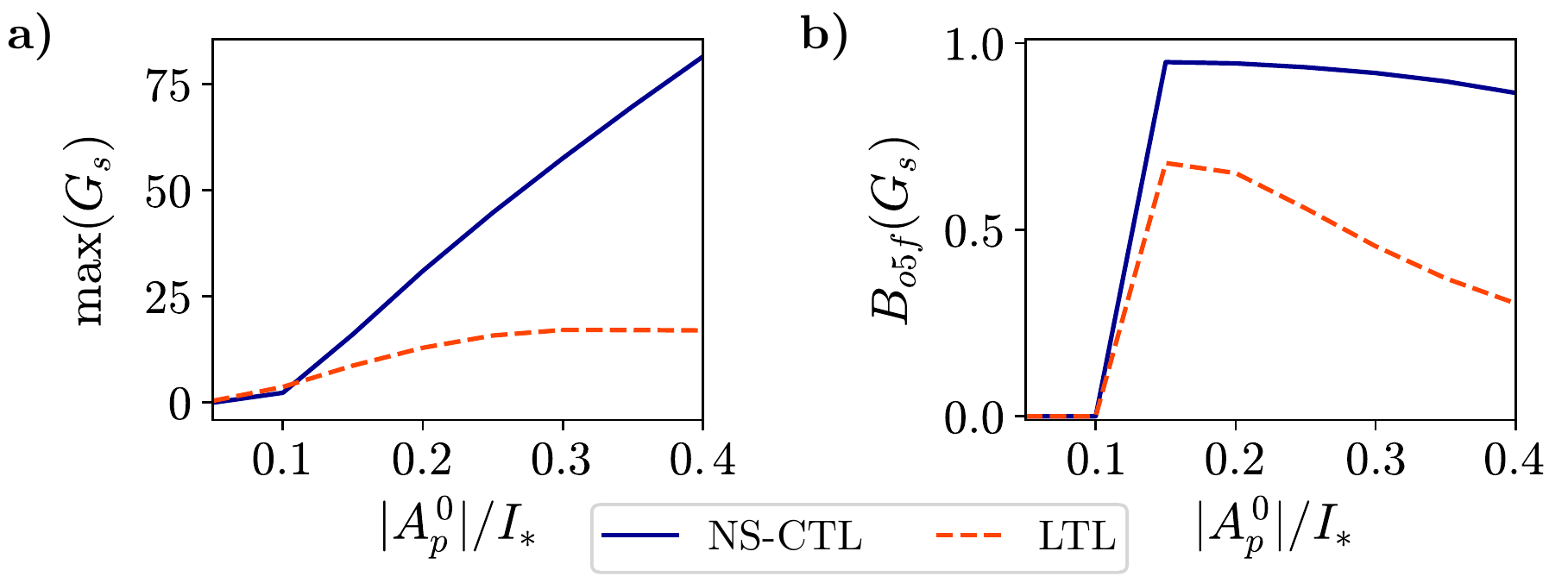}
    \caption{(a) Maximum gain and (b) over-5dB-fractional bandwidth, $B_{o5f}$, at $z/d=150$ and $\nu_p=11.63$ GHz, for various values of initial pump amplitude $A_p^0$. The maximum gain is calculated around the pump frequency, before falling to zero close to 6 GHz and 17 GHz and excluding the center zone where the gain changes curvature.}
    \label{fig:res_CPW_sb2_gains}
\end{figure}

We have also explored how $G_s$ changes by varying the magnitude of the initial pump amplitude, $\abs{A_p^0}$. To characterize this effect, we have used two figures of merit, maximum gain and fractional bandwidth for gain over 5~dB ($B_{o5f}$), plotted in \fig~\ref{fig:res_CPW_sb2_gains}. The span of $\abs{A_p^0}$ corresponds to its largest possible value so that, first, the TKIPA is still operated in the zone where the NS-CTL model is valid and, second, it is reasonably below the critical current. Both, LTL and NS-CTL models are presented.

Within the span of initial pump amplitudes, \fig~\ref{fig:res_CPW_sb2_gains} shows a stark contrast between the LTL and NS-CTL. The latter predicts larger gain with no evident saturation (panel a) and larger bandwidth (panel b).  We attribute this effect to the fact that, for the NS-CTL model, $\Theta(z)$ stabilizes over a wider range of signal frequencies, as shown in \fig~\ref{fig:res_CPW_sb2_theta_intPlot}a.

The practical result obtained from this analysis is that the NS-CTL model predicts a better performance than the LTL model if the set frequency of the pump is chosen correctly to ensure amplification.

\subsection{Design recommendations}

From the previous analyses we are able to give some recommendations for future designs of TKIPAs. Firstly, in order to improve the gain of the target signal, the pump frequency must be in the frequency zone where the NS-CTL model is valid, with the third harmonic of the pump being suppressed by a stopband. Additionally, the order of magnitude of the $RG \tilde{I}$ term must be smaller than the one of the $CL_0 \tilde{I} \omega^3/3$ term in the nonlinear wave equation for the electric current. If, furthermore, $\beta(\nu)$ is highly nonlinear around the pump frequency, large amplification of the target signal is expected. Fulfilling all these design criteria implies selecting appropriately the strength and width of the stopbands.


\section{Conclusions}
We have presented a new set of amplitude equations for TKIPAs operated at pump 
frequencies near the selected stopband (NS-CTL model). Unlike the model commonly used in the literature (LTL), the effects of complex dispersion and characteristic impedance 
have not been neglected.
We compared the two models performing simulations of a TKIPA made from a CPW Floquet transmission line. The results showed that the new NS-CTL model could predict either larger or smaller gain than the traditional one (LTL), depending on how close the pump frequency is to the stopband. This effect occurs because one of the new terms added to the amplitude equations is capable of stabilizing the phase mismatch, hence obtaining a larger gain. However, this term can also dominate the dynamics over the nonlinear term responsible for amplification since its magnitude depends on the frequency, quenching the attainable gain. Research to experimentally demonstrate these effects is underway.

\section*{Acknowledgment}

This work was partially funded by ANID through grants Fondecyt 1180700 and Basal ACE210002 and FB210003. The work of D.~Valenzuela was supported by grant DOCTORADO BECAS CHILE 2020 -- 21200705. R.~Finger acknowledges the support of FONDECYT 1221662 and FONDEF ID21I-10359.

\ifCLASSOPTIONcaptionsoff
  \newpage
\fi


\end{document}